\documentclass[prl,twocolumn,aps,showpacs,superscriptaddress]{revtex4}
\usepackage{graphicx}  % needed for figures
\usepackage{dcolumn}   % needed for some tables
\usepackage{bm}        % for math
\usepackage{amssymb}   % for math

\usepackage{amsmath}
\usepackage{braket}
\usepackage{color}

\begin{document}

\title{Elasticity of Jammed Packings of Sticky Disks}

\author{Dion J. Koeze}
\affiliation{Delft University of Technology, Process \& Energy Laboratory, Leeghwaterstraat 39, 2628 CB Delft, The Netherlands}

\author{Lingtjien Hong}
\affiliation{Delft University of Technology, Process \& Energy Laboratory, Leeghwaterstraat 39, 2628 CB Delft, The Netherlands}

\author{Abhishek Kumar}
\affiliation{Delft University of Technology, Process \& Energy Laboratory, Leeghwaterstraat 39, 2628 CB Delft, The Netherlands}

\author{Brian P. Tighe}
\affiliation{Delft University of Technology, Process \& Energy Laboratory, Leeghwaterstraat 39, 2628 CB Delft, The Netherlands}

\date{\today}

\begin{abstract}
Numerous soft materials jam into an amorphous solid at high packing fraction. This non-equilibrium phase transition is best understood in the context of a model system in which particles repel elastically when they overlap. Recently, however, it was shown that introducing any finite amount of attraction between particles changes the universality class of the transition. The properties of this new ``sticky jamming'' class remain almost entirely unexplored. We use molecular dynamics simulations and scaling analysis to determine the shear modulus, bulk modulus, and coordination of marginal solids close to the sticky jamming point. In each case, the behavior of the system departs sharply and qualitatively from the purely repulsive case. 
\end{abstract}

\maketitle

Non-Brownian dispersions such as emulsions, foams, and pastes jam into amorphous solids above a critical packing fraction \cite{liu98}. This nonequilibrium rigidity transition is best understood in a (by now canonical) model in which athermal spheres interact when they overlap \cite{durian95,ohern03}.  Nearly all prior work has focused on the case where the interaction potential is purely repulsive \cite{vanhecke10}. Nevertheless, amorphous soft matter generically displays some degree of ``stickiness,'' e.g.~due to depletion interactions in emulsions  \cite{becu06,jorjadze11,golovkova19}, finite contact angles in foams \cite{cox18}, or liquid bridges in wet granular media \cite{herminghaus05,moller07,singh14,hemmerle16}.

Amorphous solids of sticky particles differ from their repulsive counterparts: they jam at lower packing fractions, with structural signatures reminiscent of gels \cite{head07,zheng16}; they form shear bands under conditions where repulsive particles do not \cite{chaudhuri12,irani14,irani16,katgert13}; and, most tellingly, they belong to a distinct universality class \cite{lois08}. It was recently shown that repulsive jamming is a singular limit, i.e.~any finite attraction between particles places a system in the sticky jamming class \cite{koeze18}. As attraction is generic, we expect sticky jamming to be relevant to a broad range of natural and engineered systems.

What are the mechanical properties of sticky solids? 
In repulsive jamming, the elastic moduli and mean coordination display power law scaling as a function of distance to the critical packing fraction \cite{ohern03,vanhecke10}. While it seems plausible that similar scaling relations exist near sticky jamming, this hypothesis has not been tested, and the critical exponents are not known.
Here we study critical scaling in marginally jammed packings of sticky disks (Fig.~1) and show that they depart qualitatively from the familiar repulsive jamming scenario in three distinct ways. First, the shear modulus $G$  vanishes with a critical exponent that is significantly larger than its repulsive counterpart. Second, the bulk modulus $K$ also vanishes continuously at the sticky jamming point, unlike the discontinuous transition seen in repulsive systems. Finally, sticky jamming is overconstrained: at the critical packing fraction, constraints on motion (i.e.~bonds) outnumber particulate degrees of freedom. The isostatic case, where the two quantities balance precisely, is specific to repulsive jamming.

\begin{figure}[b]
\includegraphics[clip,height=0.4\linewidth]{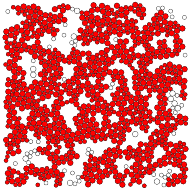} 
\hspace{0.2cm}
\centering 
\includegraphics[clip,height=0.37\linewidth]{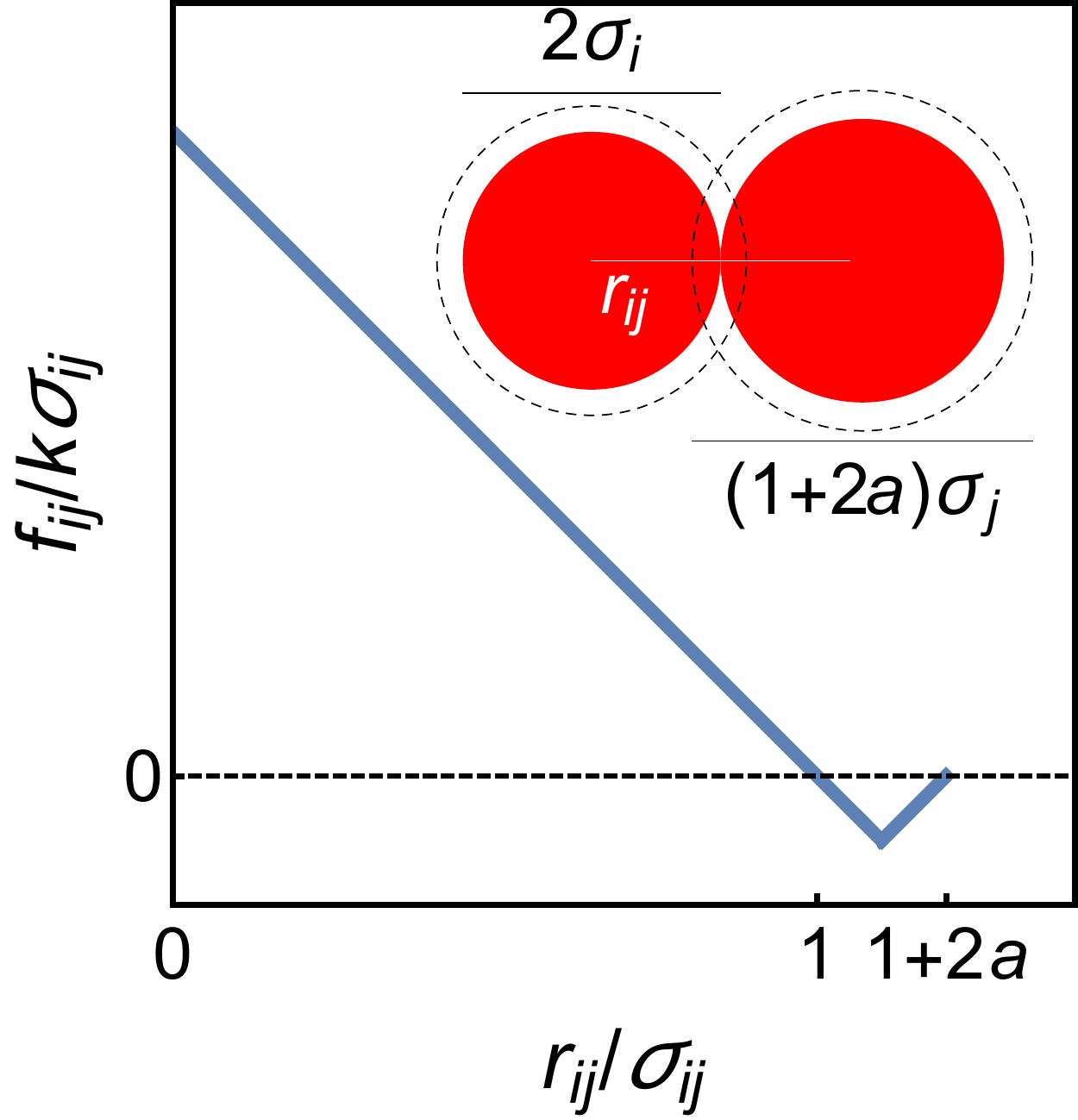} 
\caption{
(left) A periodic packing of sticky disks. Red disks participate in the spanning rigid cluster. 
(right) Their force law. The dimensionless parameter $a$ sets both the range and strength of the attractive interaction; $a = 0.1$ at left.
}
\end{figure}

{\em Model.---} We consider systems of $N = 1024$ particles in $d = 2$ dimensions prepared in a periodic square cell. Each particle has a disk-shaped core of radius $\sigma_i$ surrounded by an annular shell of thickness $a\sigma_i$. We use the standard 50:50 bidisperse mixture with a size ratio of 1.4:1 \cite{ohern03,koeze16}, and take the core diameter of the small disks as our unit of length. The finite-ranged force between disks can be expressed as a piecewise function of their overlap $\delta_{ij} = \sigma_{ij} - r_{ij}$, where $\sigma_{ij} = \sigma_{i} + \sigma_j$ and $r_{ij}$ is the distance between their centers,  
\begin{equation}
f_{ij} = \left \lbrace
\begin{array}{lc}
k\, \delta_{ij} & \delta_{ij} \ge - a \sigma_{ij} \\
-k(\delta_{ij} + 2a \sigma_{ij}) & -a \sigma_{ij} > \delta_{ij} \ge -2a\sigma_{ij}  \\
0 & \delta_{ij} < -2a\sigma_{ij} \,.
\end{array} \right.
\label{eqn:forcelaw}
\end{equation}
See also Fig.~1b. This force law is chosen both for its simplicity and for consistency with prior work \cite{head07,lois08,chaudhuri12,irani14,irani16,zheng16,koeze18}. 
Overlapping cores contribute a repulsive spring-like interaction with stiffness $k = 1$, which fixes our units of stress. Overlap between outer shells gives an attractive contribution; the parameter $a$ fixes both its range and the maximum tensile force. We use this dimensionless number to characterize attraction strength. 

Systems at a selected attraction strength and packing fraction $\phi$, calculated from the cores, are prepared by randomly placing particles in the unit cell and instantaneously quenching to a local energy minimum using a nonlinear conjugate gradient algorithm \cite{koeze16}. Each system is analyzed with the pebble game algorithm \cite{jacobs95}, which yields a complete set of rigid clusters and redundancies. 
A rigid cluster is a set of connected particles whose eigenfrequencies are all finite, apart from trivial rigid body motions. 
A redundancy is a set of bonds, any one of which can be removed from the cluster they belong to without loss of rigidity.
For each redundancy there exists a corresponding state of self-stress (SSS), a balanced configuration of forces compatible with the system's contact network. While redundancies arise naturally in the context of the pebble game, SSS's are more widely discussed in the literature -- see e.g.~\cite{calladine,alexander,roux00,tighe11b,lerner12,paulose15}.
We refer to a system as jammed if it contains a rigid cluster that spans the unit cell. 
For each jammed state the shear modulus $G$ and bulk modulus $K$ are calculated in the harmonic approximation; this requires inverting the Hessian matrix, as detailed in Refs.~\cite{maloney06,tighe11}. 

In repulsive systems the spanning cluster appears at $\phi = \phi_c(0) \approx 0.842$ \cite{ohern03,vagberg11,koeze16,koeze18}. Attractive systems jam instead at $\phi_c(a) = \phi_c(0) - \epsilon(a)$, where $\epsilon(a)$ represents the volumetric strain needed to compress the system from $\phi_c(0)$ to $\phi_c(a)$. We previously found that it scales as 
 \begin{equation}
 \epsilon(a) \simeq \left(\frac{a}{a_0}\right)^\nu \,,
 \label{eqn:epsilon}
 \end{equation}
 with $\nu \approx 0.5$ and $a_0 \approx 0.80$ \cite{koeze18}. In the same study, we found that cluster size statistics display finite size effects for attraction strengths below a characteristic scale 
$a^* \sim 1/N$, while next-nearest neighbor overlaps occur above $a \approx 0.1$. We therefore focus here on $10^{-3} \le a \le 10^{-1}$.

\begin{figure}[tb]
\centering 
\includegraphics[clip,width=0.75\linewidth]{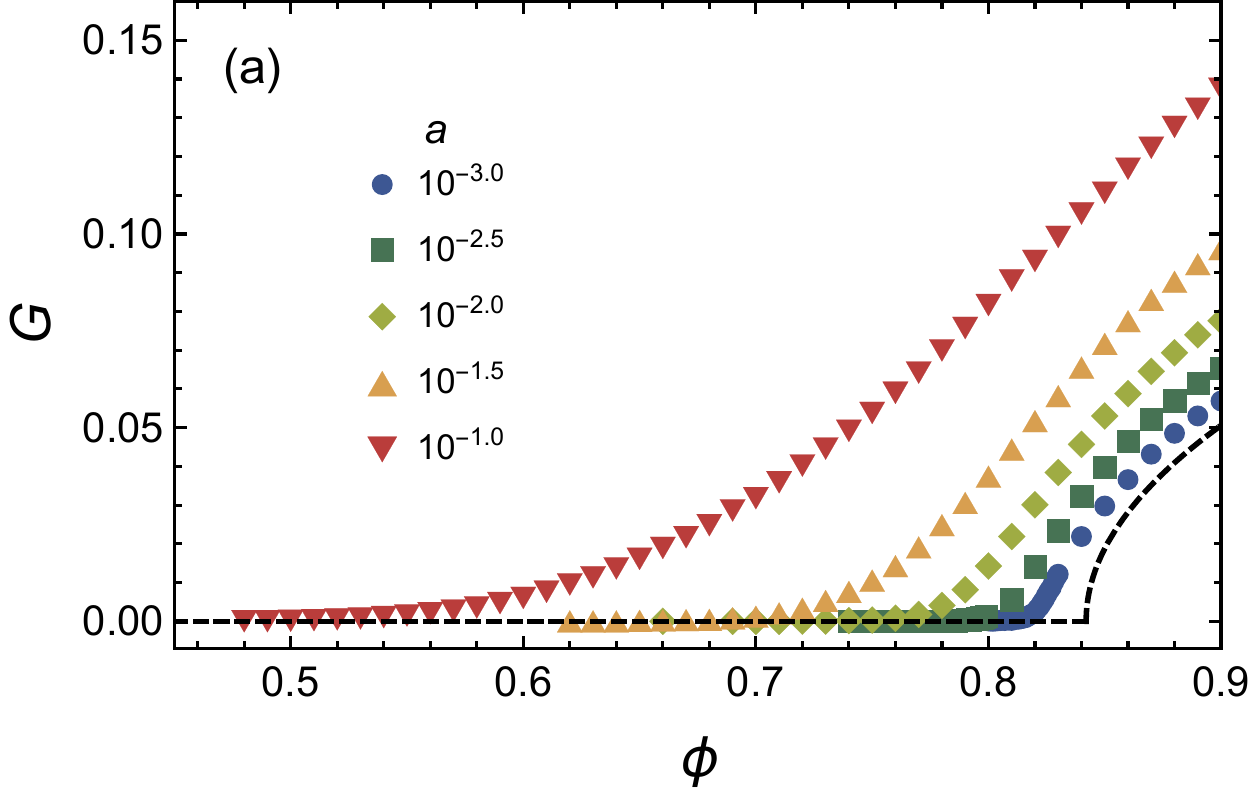} \\
\vspace{0.2cm}
\includegraphics[clip,width=0.75\linewidth]{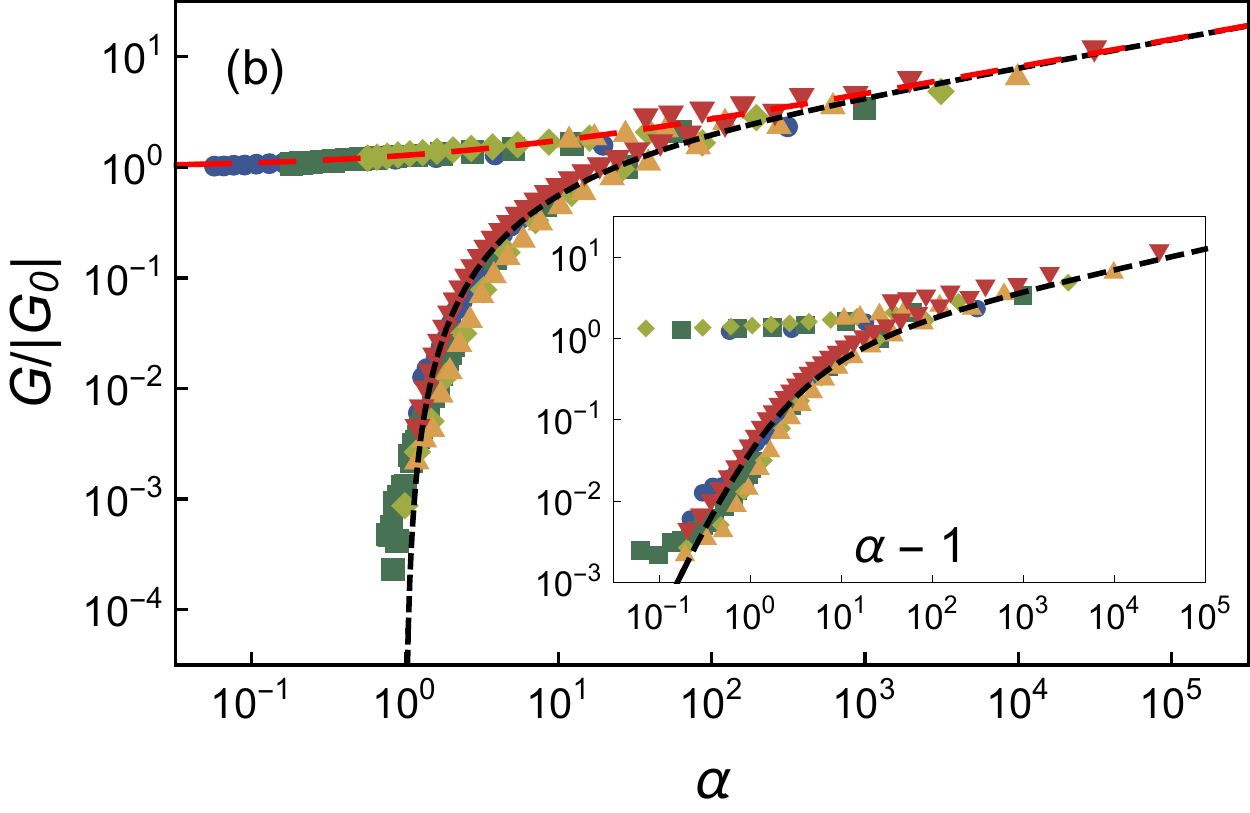}
\caption{ 
(a) The shear modulus sticky disk packings at varying attraction strengths $a$, given in the legend. This legend applies to all subsequent figures, as well.
(b) Data collapse of the rescaled shear modulus as a function of the scaling variable $\alpha$. (inset) The same data plotted versus $\alpha - 1$.}
\label{fig:G}
\end{figure}

{\em Shear modulus.---}
We first consider the shear modulus $G$ as a function of packing fraction and attraction strength. 
In Fig.~\ref{fig:G}a one sees that, while $G$ vanishes continuously for all cases, the packing fraction where it vanishes decreases with increasing $a$. We assume (and validate below) that this corresponds to the critical packing fraction determined from rigid cluster percolation. 
For the smallest attraction strengths in Fig.~1a (where $a \simeq a^*$), the modulus resembles its form in repulsive jamming, $G_0 = g_0 \, \Delta \phi_0^\mu$ with $g_0 \approx 0.22$, $\Delta \phi_0 \equiv \phi - \phi_c(0)$, and $\mu = 1/2$
\cite{ohern03}. As $a$ increases, however, the initial growth is shallower, suggestive of a power law with an exponent larger than $1$, and any resemblance to the shear modulus of a repulsive jammed state is lost.

Scaling analysis can quantify the above observations by expressing $G(\phi, a)$ in terms of a master curve $\cal G$ that depends on a single rescaled variable $\alpha$. 
To motivate $\alpha$, we note that there are three relevant packing fractions in our system: $\phi$, $\phi_c(a)$, and $\phi_c(0)$. Close to unjamming, we expect properties to scale with the distance $\phi - \phi_c(a)$, while  $\Delta \phi_0$ will become relevant deeper in the jammed phase. One could therefore construct $\alpha$ from the ratio of these two distances. 
An equivalent (and ultimately more practical) approach is to replace the distance to the sticky jamming point with the distance between the repulsive and sticky jamming points, $\epsilon(a)$, i.e.~to choose 
\begin{equation}
\alpha \equiv \left( \frac{\epsilon}{|\Delta \phi_0|} \right)^{1/\nu} \simeq \frac{a/a_0}{|\Delta \phi_0|^{1/\nu}} \,.
\end{equation}
Note that $\alpha$ has been defined so as to be linear in $a$. We then make the scaling ansatz
\begin{equation}
{G} = {|G_0|} \, {\cal G}_\pm(\alpha)\,.
\label{eqn:Gansatz}
\end{equation}
The function ${\cal G}_\pm(\alpha)$ has one branch for each sign of $\Delta \phi_0$.  

A test of the scaling ansatz is shown in Fig.~\ref{fig:G}b. There is indeed good data collapse.
The upper branch ${\cal G}_+$ approaches unity when $\alpha$ vanishes, i.e.~the modulus for repulsive jamming is recovered. The lower branch ${\cal G}_-$ vanishes for states below $\alpha \approx 1$, indicative of unjamming. 
Some states do exist slightly below $\alpha = 1$; we attribute this to finite size effects, which smear out $\phi_c(a)$ \cite{koeze18}.
It is apparent that there is also a third scaling regime, in which both branches scale as ${\cal G}_\pm \sim \alpha^\Delta$ for some positive exponent $\Delta$.
This expression describes the shear modulus when 
$\phi \approx \phi_c(0)$.
Since $G$ remains finite at $\phi_c(0)$ when $a$ is nonzero, 
any dependence on $\Delta \phi_0$ must be subdominant. This requirement is only satisfied if $\Delta = \mu \nu = 0.25$, which we verify below. 

Clearly  the initial growth of $G$ in sticky jammed systems does not follow the square root scaling seen in repulsive jamming. 
This observation can be quantified with the aid of Eq.~(\ref{eqn:Gansatz}). 
We begin by making the ansatz that, sufficiently close to $\phi_c(a)$, $G$ grows as a power law, 
\begin{equation}
G \simeq  f(a) \, [\phi -  \phi_c(a)]^\psi \,,
\label{eqn:scalingfunction}
\end{equation}
for some exponent $\psi > 0$.
As there is no reason to forbid it, Eq.~(\ref{eqn:scalingfunction}) includes a prefactor $f(a)$ that depends on attraction strength. We assume $f \simeq c_G \, a^{-\omega}$ to leading order. Importantly, the exponent $\omega$ is not free; in order to be compatible with the scaling ansatz (\ref{eqn:Gansatz}), it must be $\omega = \nu \psi - \Delta$.
Hence Eq.~(\ref{eqn:scalingfunction}) can be re-written as 
\begin{equation}
{\cal G}_- \simeq c_G \,  \alpha^\Delta \left[1 - \alpha^{-\nu}\right]^\psi \,.% \,\,\,\,{\rm as} \,\, \alpha \searrow 1 \,.
\label{eqn:calGalpha}
\end{equation}
Eq.~(\ref{eqn:calGalpha}) provides an excellent fit to the data in Fig.~\ref{fig:G}a for $\psi = 2.5$ (hence $\omega = 1.0$) and $c_G = 0.81$. It correctly captures the full form of the lower branch ${\cal G}_-$, including the asymptotic limits $\alpha \rightarrow 1^+$ (see also the inset to Fig.~\ref{fig:G}b, which plots the same data versus $\alpha - 1$), and $\alpha \rightarrow \infty$ (validating $\Delta = 0.25$). 
We conclude  that the shear modulus in sticky disk packings is well described by Eq.~(\ref{eqn:scalingfunction}) for $\phi_c(a) \le \phi \lesssim \phi_c(0)$. 
This (re-)confirms that sticky jammed states have a different mechanical character from their repulsive counterparts.

To complete our description of the scaling function, 
we note that the form ${\cal G}_+ = [1 + (c_G \,  \alpha^\Delta)^2]^{1/2}$ 
has the required asymptotics and fits the upper branch  over the full range of $\alpha$ -- see the the long-dashed curve in Fig.~\ref{fig:G}b.

{\em Bulk modulus.---}
\begin{figure}[tbh]
\centering 
\includegraphics[clip,width=0.75\linewidth]{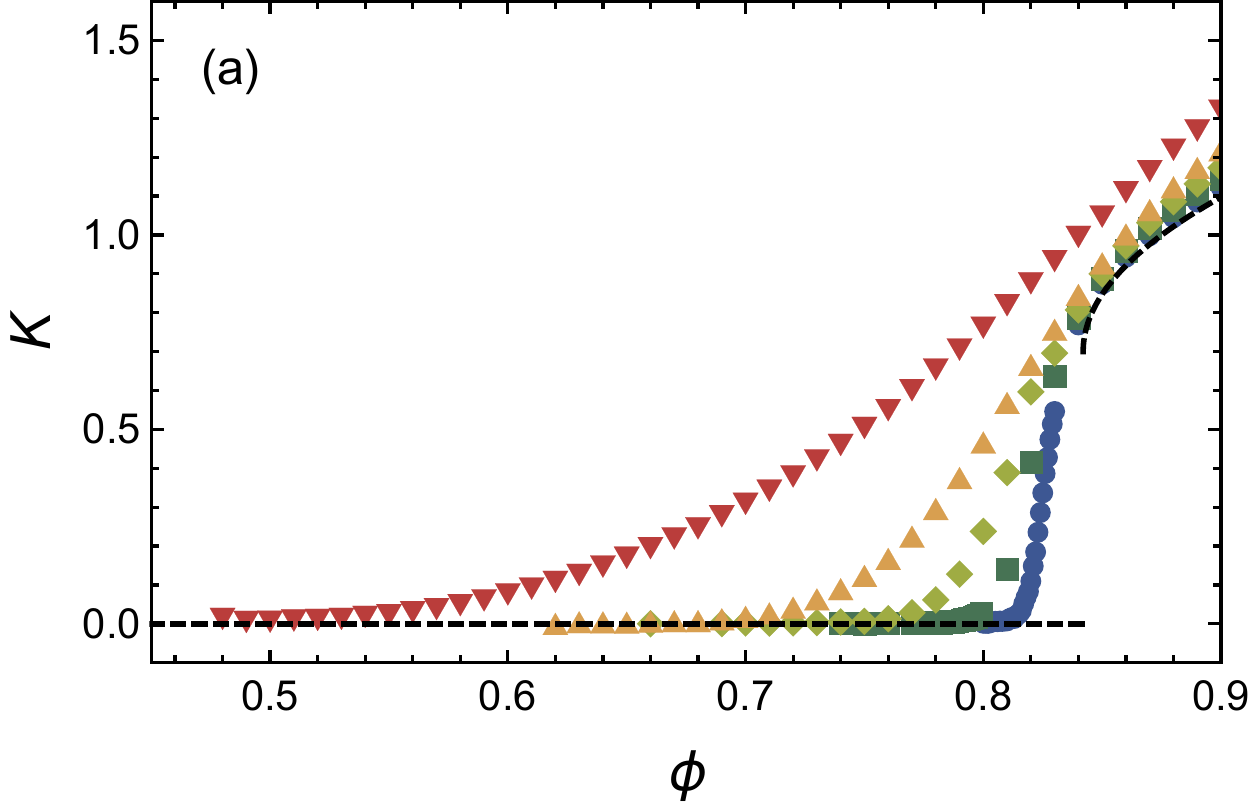} \\
\vspace{0.2cm}
\includegraphics[clip,width=0.758\linewidth]{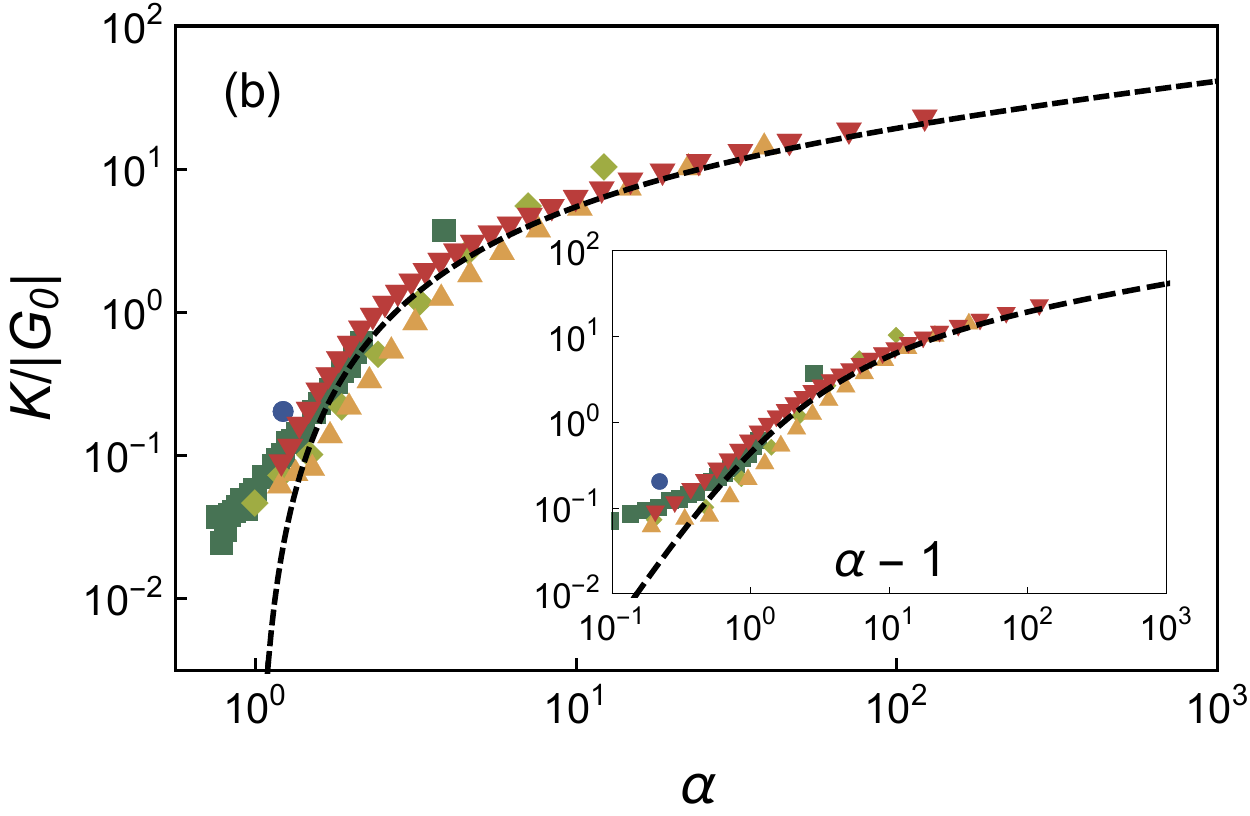} \\
\caption{
(a) The bulk modulus of sticky disk packings at varying attraction strengths $a$.
(b) The rescaled bulk modulus for $\phi \le 0.81$ plotted as a function of the scaling variable $\alpha$ (main panel) and $\alpha - 1$ (inset). }
\label{fig:K}
\end{figure}
Next we consider the bulk modulus $K$ as a function of packing fraction for varying attraction strength, as shown in Fig.~\ref{fig:K}a.  
Note that there is an important difference between $G$ and $K$ in repulsive systems -- while  the shear modulus vanishes continuously at $\phi_c(0)$, the bulk modulus undergoes a jump (dashed curve). In systems with finite attraction strength, however, the bulk modulus is continuous, and grows in a manner reminiscent of the shear modulus.
 
Because of the discontinuity in the repulsive bulk modulus, it is not possible to collapse the data of Fig.~\ref{fig:K}a to a two-branched master curve analogous to Fig.~\ref{fig:G}b. 
However, the fact that $K$ in sticky systems vanishes continuously suggests that our approach to $G$ may also be effective in describing the bulk modulus when the packing fraction is sufficiently far below  $\phi_c(0)$. In Fig.~\ref{fig:K}b we plot 
$K/|G_0|$ versus $\alpha$, restricted to packings where $\phi \le 0.81$. 
We find that the collapsed data are well described by the function ${\cal K}_- = 9.8 \, {\cal G}_-$. We conclude that the same critical exponents govern both moduli close to $\phi_c(a)$.

\begin{figure}[tb]
\centering 
\includegraphics[clip,width=0.75\linewidth]{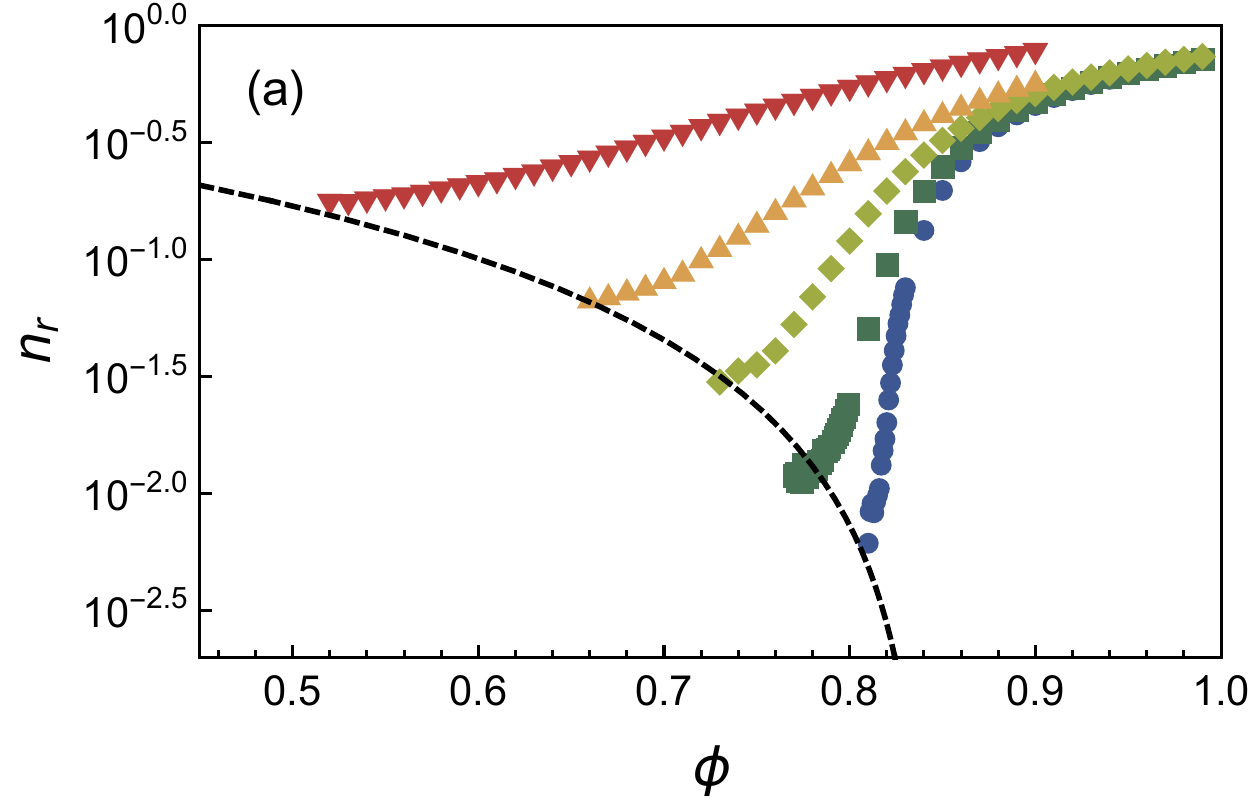}\\
\vspace{0.2cm}
\includegraphics[clip,width=0.75\linewidth]{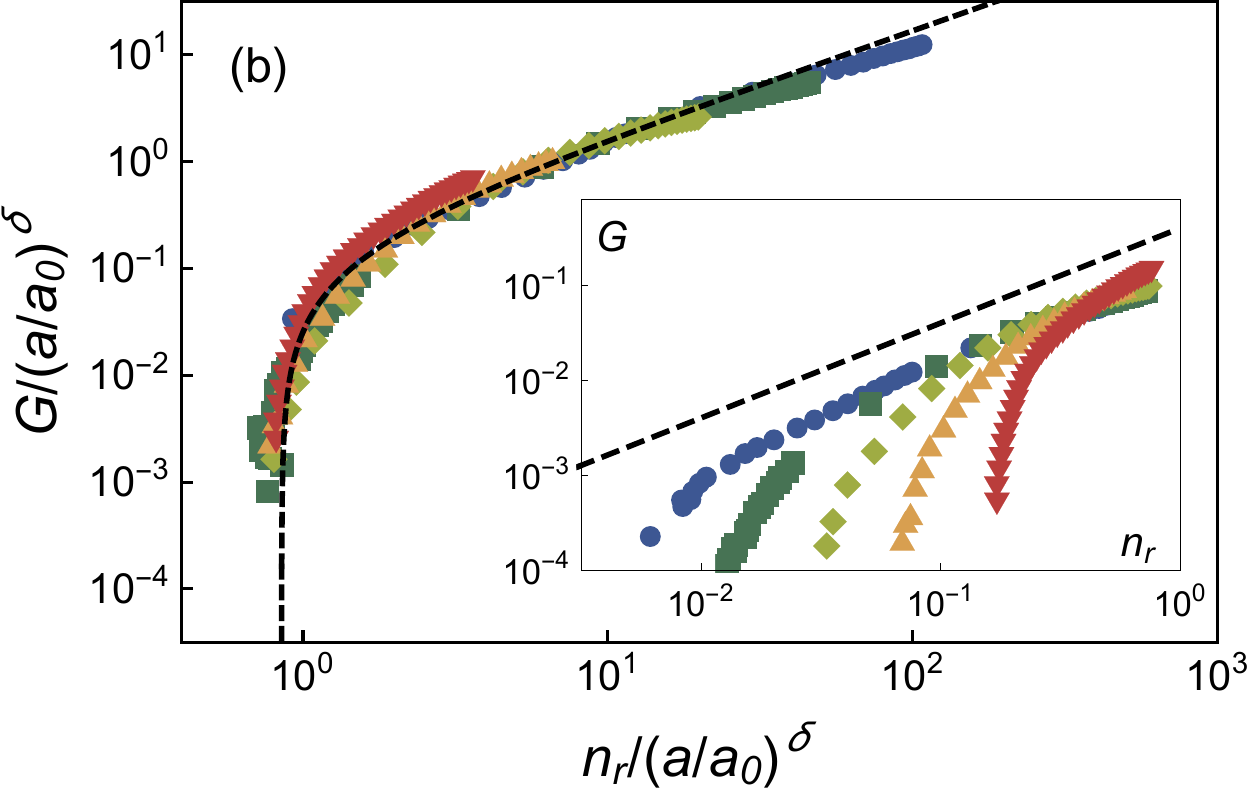} 
\caption{(a) The redundancy density $n_r$ plotted as a function of packing fraction for varying attraction strengths $a$. 
(b) Collapse of the shear modulus with $n_r$ when both are rescaled with $a^\delta$. (inset) The same data without rescaling. The dashed line has slope 1.}
\label{fig:nr}
\end{figure}

{\em Coordination and redundancies.---}
The mean coordination $z$ characterizes contact network structure and plays a fundamental role in theoretical predictions for the moduli of repulsive jammed packings, e.g.~\cite{wyartannales,schlegel16}.  It is therefore useful to investigate the relationship between coordination and packing fraction. 

Let us first recall the main result of Maxwell-Calladine counting, which relates degrees of freedom, constraints on motion, floppy modes, and redundancies in a network of nodes and bonds (viz.~particles and contacts) \cite{calladine}. It states that $d - \frac{1}{2} z = n_{f} - n_r + O(1/N)$,
where $n_f$ and $n_r$ are the numbers of floppy modes and redundancies per node, respectively. We will neglect the $O(1/N)$ correction, which depends on boundary conditions. 
Applying this relation to a spanning rigid cluster, which has no floppy modes, gives the mean coordination $z = z_{\rm iso} + 2n_r$, where $z_{\rm iso} \equiv 2d$ is the Maxwell isostatic value. Therefore compressing the system, which creates new contacts, also introduces an equal number of redundancies. 
While Maxwell-Calladine counting places no further constraints on $n_r$, it is an empirical fact that $n_r$ vanishes at the jamming point in repulsive systems
\cite{ohern03,goodrich12,dagois-bohy12,goodrich14}, i.e.~$z_c(a = 0) = z_{\rm iso}$. In fact, $z = z_{\rm iso}$ is often used as a criterion for jamming.  Given this context, in Fig.~\ref{fig:nr}a we plot $n_r$ as a function of $\phi$ for varying $a$. Unlike repulsive systems, $n_r$ remains finite as $\phi$ approaches $\phi_c(a)$, indicating that the spanning cluster is overconstrained, $z_c(a) > z_{\rm iso}$. 
Hence the Maxwell isostatic value does not signal rigidity percolation in sticky systems.

In order to quantify the critical redundancy density $n_c(a) \equiv n_r(\phi_c(a))$, we investigate the point where the shear modulus vanishes in a plot of $G$ versus $n_r$, as shown in the inset of Fig.~\ref{fig:nr}b. We have verified that the bulk modulus also vanishes at the same redundancy density (not shown). We seek to collapse the data by plotting $(a/a_0)^\delta G$ versus $n_r/(a/a_0)^\delta$; the same exponent $\delta$  must appear on each axis to ensure that $G \sim n_r$, the known form for repulsive systems \cite{ohern03,wyartannales},  is recovered when when $a \rightarrow 0$. We find good data collapse for $\delta = 0.75$, with the master curve vanishing at a value $n_0 \approx 0.85$ (Fig.~\ref{fig:nr}b, main panel). This implies that $n_c \simeq  n_0 \, (a/a_0)^\delta$, and that the excess coordination at the attractive jamming point is $z_c(a) - z_{\rm iso} = 2n_c \sim [\phi_c(0) - \phi_c(a)]^{\delta /\nu}$ (see Fig.~\ref{fig:nr}a, dashed curve). 
Above the sticky jamming point, the expression $G \sim n_r - n_c$ represents a natural generalization of the repulsive case and, further, provides a reasonable fit to our data (dashed curve, Fig.~\ref{fig:nr}b main panel). In other words, $G$ grows in proportion to the number of redundancies in excess of those present in the spanning cluster at percolation.

{\em Discussion.---} 
We have shown that sticky jamming differs from repulsive jamming in three distinct ways. While the shear modulus in repulsive packings vanishes continuously with a critical exponent $\mu =1/2$, in sticky jamming the exponent $\psi \approx 2.5$ is much larger. The bulk modulus in sticky systems vanishes continuously and in proportion to $G$, unlike repulsive jamming where it is discontinuous. And redundancies persist at the sticky jamming point, with number density $n_c \sim a^\delta$ and $\delta \approx 0.75$. In contrast, $n_c$ vanishes at the repulsive jamming point. 

The mechanical and structural properties identified here represent  a challenge to existing theories of elasticity in marginal solids \cite{feng85,wyartannales,schlegel16,merkel19}.
A successful theory should predict the values of the exponents $\psi$, $\delta$, and $\nu$, 
each of which remain empirical.
More qualitatively, we are not aware of any theory that predicts unjamming with $n_c > 0$ (and hence $z > z_{\rm iso}$). As $z_{\rm iso} = 2d$ results from a mean field counting argument, explaining this result may require a non-mean field theory. 
Effective medium theories for marginal elastic solids predict both $G$ and $K$ to vanish continuously  \cite{feng85}; however, they do not successfully account for the repulsive case \cite{ellenbroek09b}.
While there is a successful theory of elasticity in repulsive jammed solids \cite{wyartannales}, a straightforward generalization of its results would predict both $G$ and $K$ are {\em dis}continuous at sticky unjamming due to the presence of redundancies.  

There are several obvious directions for future work. 
The force law in Eq.~(\ref{eqn:forcelaw}) is particularly simple, with just one parameter $a$; untangling the role of, e.g., the maximum tensile force and the range of the interaction will facilitate comparisons to experiment. 
Finite size scaling analysis will help to validate our conclusions and to sharpen estimated values for the critical exponents. 
While the results here are for $d = 2$ dimensions,  we expect to find critical scaling in three dimensions as well. However, the critical exponents  $\psi$, $\delta$, and $\nu$ may differ, as is the case for other exponents near sticky jamming \cite{lois08}. 
And finally, it is natural to ask how sticky systems respond under small- and large-amplitude oscillatory shear, which would build a bridge between the present work and viscoelasticity in repulsive jamming  \cite{tighe11,baumgarten17b,dagois-bohy17}, as well as steady shear flow in sticky systems \cite{irani14,irani16}.

We acknowledge financial support from the Netherlands Organization for Scientific Research (NWO). This work was sponsored by NWO Physical Sciences through the use of supercomputer facilities.

\bibliographystyle{apsrev}
%\bibliography{tighe_bib}   % name your BibTeX data base

\begin{thebibliography}{41}
\expandafter\ifx\csname natexlab\endcsname\relax\def\natexlab#1{#1}\fi
\expandafter\ifx\csname bibnamefont\endcsname\relax
  \def\bibnamefont#1{#1}\fi
\expandafter\ifx\csname bibfnamefont\endcsname\relax
  \def\bibfnamefont#1{#1}\fi
\expandafter\ifx\csname citenamefont\endcsname\relax
  \def\citenamefont#1{#1}\fi
\expandafter\ifx\csname url\endcsname\relax
  \def\url#1{\texttt{#1}}\fi
\expandafter\ifx\csname urlprefix\endcsname\relax\def\urlprefix{URL }\fi
\providecommand{\bibinfo}[2]{#2}
\providecommand{\eprint}[2][]{\url{#2}}

\bibitem[{\citenamefont{Liu and Nagel}(1998)}]{liu98}
\bibinfo{author}{\bibfnamefont{A.~J.} \bibnamefont{Liu}} \bibnamefont{and}
  \bibinfo{author}{\bibfnamefont{S.~R.} \bibnamefont{Nagel}},
  \bibinfo{journal}{Nature} \textbf{\bibinfo{volume}{396}}, \bibinfo{pages}{21}
  (\bibinfo{year}{1998}).

\bibitem[{\citenamefont{Durian}(1995)}]{durian95}
\bibinfo{author}{\bibfnamefont{D.~J.} \bibnamefont{Durian}},
  \bibinfo{journal}{Phys. Rev. Lett.} \textbf{\bibinfo{volume}{75}},
  \bibinfo{pages}{4780} (\bibinfo{year}{1995}).

\bibitem[{\citenamefont{O'Hern et~al.}(2003)\citenamefont{O'Hern, Silbert, Liu,
  and Nagel}}]{ohern03}
\bibinfo{author}{\bibfnamefont{C.~S.} \bibnamefont{O'Hern}},
  \bibinfo{author}{\bibfnamefont{L.~E.} \bibnamefont{Silbert}},
  \bibinfo{author}{\bibfnamefont{A.~J.} \bibnamefont{Liu}}, \bibnamefont{and}
  \bibinfo{author}{\bibfnamefont{S.~R.} \bibnamefont{Nagel}},
  \bibinfo{journal}{Phys.~Rev.~E} \textbf{\bibinfo{volume}{68}},
  \bibinfo{pages}{011306} (\bibinfo{year}{2003}).

\bibitem[{\citenamefont{van Hecke}(2010)}]{vanhecke10}
\bibinfo{author}{\bibfnamefont{M.}~\bibnamefont{van Hecke}},
  \bibinfo{journal}{J.~Phys.~Cond.~Matt.} \textbf{\bibinfo{volume}{22}},
  \bibinfo{pages}{033101} (\bibinfo{year}{2010}).

\bibitem[{\citenamefont{B{\'e}cu et~al.}(2006)\citenamefont{B{\'e}cu,
  Manneville, and Colin}}]{becu06}
\bibinfo{author}{\bibfnamefont{L.}~\bibnamefont{B{\'e}cu}},
  \bibinfo{author}{\bibfnamefont{S.}~\bibnamefont{Manneville}},
  \bibnamefont{and} \bibinfo{author}{\bibfnamefont{A.}~\bibnamefont{Colin}},
  \bibinfo{journal}{Phys. Rev. Lett.} \textbf{\bibinfo{volume}{96}},
  \bibinfo{pages}{138302} (\bibinfo{year}{2006}).

\bibitem[{\citenamefont{Jorjadze et~al.}(2011)\citenamefont{Jorjadze, Pontani,
  Newhall, and Bruji{\'c}}}]{jorjadze11}
\bibinfo{author}{\bibfnamefont{I.}~\bibnamefont{Jorjadze}},
  \bibinfo{author}{\bibfnamefont{L.-L.} \bibnamefont{Pontani}},
  \bibinfo{author}{\bibfnamefont{K.~A.} \bibnamefont{Newhall}},
  \bibnamefont{and}
  \bibinfo{author}{\bibfnamefont{J.}~\bibnamefont{Bruji{\'c}}},
  \bibinfo{journal}{Proc. Nat. Acad. Sci.} \textbf{\bibinfo{volume}{108}},
  \bibinfo{pages}{4286} (\bibinfo{year}{2011}).

\bibitem[{\citenamefont{Golovkova et~al.}(2019)\citenamefont{Golovkova, Montel,
  Wandersman, Bertrand, Prevost, and Pontani}}]{golovkova19}
\bibinfo{author}{\bibfnamefont{I.}~\bibnamefont{Golovkova}},
  \bibinfo{author}{\bibfnamefont{L.}~\bibnamefont{Montel}},
  \bibinfo{author}{\bibfnamefont{E.}~\bibnamefont{Wandersman}},
  \bibinfo{author}{\bibfnamefont{T.}~\bibnamefont{Bertrand}},
  \bibinfo{author}{\bibfnamefont{A.~M.} \bibnamefont{Prevost}},
  \bibnamefont{and} \bibinfo{author}{\bibfnamefont{L.-L.}
  \bibnamefont{Pontani}} (\bibinfo{year}{2019}), \eprint{arXiv:1911.12699}.

\bibitem[{\citenamefont{Cox et~al.}(2018)\citenamefont{Cox, Kraynik, Weaire,
  and Hutzler}}]{cox18}
\bibinfo{author}{\bibfnamefont{S.}~\bibnamefont{Cox}},
  \bibinfo{author}{\bibfnamefont{A.}~\bibnamefont{Kraynik}},
  \bibinfo{author}{\bibfnamefont{D.}~\bibnamefont{Weaire}}, \bibnamefont{and}
  \bibinfo{author}{\bibfnamefont{S.}~\bibnamefont{Hutzler}},
  \bibinfo{journal}{Soft Matter} \textbf{\bibinfo{volume}{14}},
  \bibinfo{pages}{5922} (\bibinfo{year}{2018}).

\bibitem[{\citenamefont{Herminghaus}(2005)}]{herminghaus05}
\bibinfo{author}{\bibfnamefont{S.}~\bibnamefont{Herminghaus}},
  \bibinfo{journal}{Advances in Physics} \textbf{\bibinfo{volume}{54}},
  \bibinfo{pages}{221} (\bibinfo{year}{2005}).

\bibitem[{\citenamefont{M{\o}ller and Bonn}(2007)}]{moller07}
\bibinfo{author}{\bibfnamefont{P.~C.} \bibnamefont{M{\o}ller}}
  \bibnamefont{and} \bibinfo{author}{\bibfnamefont{D.}~\bibnamefont{Bonn}},
  \bibinfo{journal}{Europhys.~Lett.} \textbf{\bibinfo{volume}{80}},
  \bibinfo{pages}{38002} (\bibinfo{year}{2007}).

\bibitem[{\citenamefont{Singh et~al.}(2014)\citenamefont{Singh, Magnanimo,
  Saitoh, and Luding}}]{singh14}
\bibinfo{author}{\bibfnamefont{A.}~\bibnamefont{Singh}},
  \bibinfo{author}{\bibfnamefont{V.}~\bibnamefont{Magnanimo}},
  \bibinfo{author}{\bibfnamefont{K.}~\bibnamefont{Saitoh}}, \bibnamefont{and}
  \bibinfo{author}{\bibfnamefont{S.}~\bibnamefont{Luding}},
  \bibinfo{journal}{Phys. Rev. E} \textbf{\bibinfo{volume}{90}},
  \bibinfo{pages}{022202} (\bibinfo{year}{2014}).

\bibitem[{\citenamefont{Hemmerle et~al.}(2016)\citenamefont{Hemmerle,
  Schr{\"o}ter, and Goehring}}]{hemmerle16}
\bibinfo{author}{\bibfnamefont{A.}~\bibnamefont{Hemmerle}},
  \bibinfo{author}{\bibfnamefont{M.}~\bibnamefont{Schr{\"o}ter}},
  \bibnamefont{and} \bibinfo{author}{\bibfnamefont{L.}~\bibnamefont{Goehring}},
  \bibinfo{journal}{Scientific Reports} \textbf{\bibinfo{volume}{6}},
  \bibinfo{pages}{35650} (\bibinfo{year}{2016}).

\bibitem[{\citenamefont{Head}(2007)}]{head07}
\bibinfo{author}{\bibfnamefont{D.}~\bibnamefont{Head}}, \bibinfo{journal}{The
  European Physical Journal E} \textbf{\bibinfo{volume}{22}},
  \bibinfo{pages}{151} (\bibinfo{year}{2007}).

\bibitem[{\citenamefont{Zheng et~al.}(2016)\citenamefont{Zheng, Liu, and
  Xu}}]{zheng16}
\bibinfo{author}{\bibfnamefont{W.}~\bibnamefont{Zheng}},
  \bibinfo{author}{\bibfnamefont{H.}~\bibnamefont{Liu}}, \bibnamefont{and}
  \bibinfo{author}{\bibfnamefont{N.}~\bibnamefont{Xu}}, \bibinfo{journal}{Phys.
  Rev. E} \textbf{\bibinfo{volume}{94}}, \bibinfo{pages}{062608}
  (\bibinfo{year}{2016}).

\bibitem[{\citenamefont{Chaudhuri et~al.}(2012)\citenamefont{Chaudhuri,
  Berthier, and Bocquet}}]{chaudhuri12}
\bibinfo{author}{\bibfnamefont{P.}~\bibnamefont{Chaudhuri}},
  \bibinfo{author}{\bibfnamefont{L.}~\bibnamefont{Berthier}}, \bibnamefont{and}
  \bibinfo{author}{\bibfnamefont{L.}~\bibnamefont{Bocquet}},
  \bibinfo{journal}{Phys. Rev. E} \textbf{\bibinfo{volume}{85}},
  \bibinfo{pages}{021503} (\bibinfo{year}{2012}).

\bibitem[{\citenamefont{Irani et~al.}(2014)\citenamefont{Irani, Chaudhuri, and
  Heussinger}}]{irani14}
\bibinfo{author}{\bibfnamefont{E.}~\bibnamefont{Irani}},
  \bibinfo{author}{\bibfnamefont{P.}~\bibnamefont{Chaudhuri}},
  \bibnamefont{and}
  \bibinfo{author}{\bibfnamefont{C.}~\bibnamefont{Heussinger}},
  \bibinfo{journal}{Phys. Rev. Lett.} \textbf{\bibinfo{volume}{112}},
  \bibinfo{pages}{188303} (\bibinfo{year}{2014}).

\bibitem[{\citenamefont{Irani et~al.}(2016)\citenamefont{Irani, Chaudhuri, and
  Heussinger}}]{irani16}
\bibinfo{author}{\bibfnamefont{E.}~\bibnamefont{Irani}},
  \bibinfo{author}{\bibfnamefont{P.}~\bibnamefont{Chaudhuri}},
  \bibnamefont{and}
  \bibinfo{author}{\bibfnamefont{C.}~\bibnamefont{Heussinger}},
  \bibinfo{journal}{Phys. Rev. E} \textbf{\bibinfo{volume}{94}},
  \bibinfo{pages}{052608} (\bibinfo{year}{2016}).

\bibitem[{\citenamefont{Katgert et~al.}(2013)\citenamefont{Katgert, Tighe, and
  van Hecke}}]{katgert13}
\bibinfo{author}{\bibfnamefont{G.}~\bibnamefont{Katgert}},
  \bibinfo{author}{\bibfnamefont{B.~P.} \bibnamefont{Tighe}}, \bibnamefont{and}
  \bibinfo{author}{\bibfnamefont{M.}~\bibnamefont{van Hecke}},
  \bibinfo{journal}{Soft Matter} \textbf{\bibinfo{volume}{9}},
  \bibinfo{pages}{9739} (\bibinfo{year}{2013}).

\bibitem[{\citenamefont{Lois et~al.}(2008)\citenamefont{Lois, Blawzdziewicz,
  and O'Hern}}]{lois08}
\bibinfo{author}{\bibfnamefont{G.}~\bibnamefont{Lois}},
  \bibinfo{author}{\bibfnamefont{J.}~\bibnamefont{Blawzdziewicz}},
  \bibnamefont{and} \bibinfo{author}{\bibfnamefont{C.~S.}
  \bibnamefont{O'Hern}}, \bibinfo{journal}{Phys. Rev. Lett.}
  \textbf{\bibinfo{volume}{100}}, \bibinfo{pages}{028001}
  (\bibinfo{year}{2008}).

\bibitem[{\citenamefont{Koeze and Tighe}(2018)}]{koeze18}
\bibinfo{author}{\bibfnamefont{D.~J.} \bibnamefont{Koeze}} \bibnamefont{and}
  \bibinfo{author}{\bibfnamefont{B.~P.} \bibnamefont{Tighe}},
  \bibinfo{journal}{Phys. Rev. Lett.} \textbf{\bibinfo{volume}{121}},
  \bibinfo{pages}{188002} (\bibinfo{year}{2018}).

\bibitem[{\citenamefont{Koeze et~al.}(2016)\citenamefont{Koeze, V{\aa}gberg,
  Tjoa, and Tighe}}]{koeze16}
\bibinfo{author}{\bibfnamefont{D.~J.} \bibnamefont{Koeze}},
  \bibinfo{author}{\bibfnamefont{D.}~\bibnamefont{V{\aa}gberg}},
  \bibinfo{author}{\bibfnamefont{B.~B.} \bibnamefont{Tjoa}}, \bibnamefont{and}
  \bibinfo{author}{\bibfnamefont{B.~P.} \bibnamefont{Tighe}},
  \bibinfo{journal}{EPL} \textbf{\bibinfo{volume}{113}}, \bibinfo{pages}{54001}
  (\bibinfo{year}{2016}).

\bibitem[{\citenamefont{Jacobs and Thorpe}(1995)}]{jacobs95}
\bibinfo{author}{\bibfnamefont{D.~J.} \bibnamefont{Jacobs}} \bibnamefont{and}
  \bibinfo{author}{\bibfnamefont{M.~F.} \bibnamefont{Thorpe}},
  \bibinfo{journal}{Phys. Rev. Lett.} \textbf{\bibinfo{volume}{75}},
  \bibinfo{pages}{4051} (\bibinfo{year}{1995}).

\bibitem[{\citenamefont{Pellegrino and Calladine}(1986)}]{calladine}
\bibinfo{author}{\bibfnamefont{S.}~\bibnamefont{Pellegrino}} \bibnamefont{and}
  \bibinfo{author}{\bibfnamefont{C.~R.} \bibnamefont{Calladine}},
  \bibinfo{journal}{Int.~J.~Solids Structures} \textbf{\bibinfo{volume}{22}},
  \bibinfo{pages}{409} (\bibinfo{year}{1986}).

\bibitem[{\citenamefont{Alexander}(1998)}]{alexander}
\bibinfo{author}{\bibfnamefont{S.}~\bibnamefont{Alexander}},
  \bibinfo{journal}{Phys.~Rep} \textbf{\bibinfo{volume}{296}},
  \bibinfo{pages}{65} (\bibinfo{year}{1998}).

\bibitem[{\citenamefont{Roux}(2000)}]{roux00}
\bibinfo{author}{\bibfnamefont{J.-N.} \bibnamefont{Roux}},
  \bibinfo{journal}{Phys. Rev. E} \textbf{\bibinfo{volume}{61}},
  \bibinfo{pages}{6802} (\bibinfo{year}{2000}).

\bibitem[{\citenamefont{Tighe and Vlugt}(2011)}]{tighe11b}
\bibinfo{author}{\bibfnamefont{B.~P.} \bibnamefont{Tighe}} \bibnamefont{and}
  \bibinfo{author}{\bibfnamefont{T.~J.~H.} \bibnamefont{Vlugt}},
  \bibinfo{journal}{Journal of Statistical Mechanics: Theory and Experiment} p.
  \bibinfo{pages}{P04002} (\bibinfo{year}{2011}).

\bibitem[{\citenamefont{Lerner et~al.}(2012)\citenamefont{Lerner, D\"uring, and
  Wyart}}]{lerner12}
\bibinfo{author}{\bibfnamefont{E.}~\bibnamefont{Lerner}},
  \bibinfo{author}{\bibfnamefont{G.}~\bibnamefont{D\"uring}}, \bibnamefont{and}
  \bibinfo{author}{\bibfnamefont{M.}~\bibnamefont{Wyart}},
  \bibinfo{journal}{Proc. Nat. Acad. Sci. (USA)}
  \textbf{\bibinfo{volume}{109}}, \bibinfo{pages}{4798} (\bibinfo{year}{2012}).

\bibitem[{\citenamefont{Paulose et~al.}(2015)\citenamefont{Paulose, Meeussen,
  and Vitelli}}]{paulose15}
\bibinfo{author}{\bibfnamefont{J.}~\bibnamefont{Paulose}},
  \bibinfo{author}{\bibfnamefont{A.~S.} \bibnamefont{Meeussen}},
  \bibnamefont{and} \bibinfo{author}{\bibfnamefont{V.}~\bibnamefont{Vitelli}},
  \bibinfo{journal}{Proc. Nat. Acad. Sci. (USA)}
  \textbf{\bibinfo{volume}{112}}, \bibinfo{pages}{7639} (\bibinfo{year}{2015}).

\bibitem[{\citenamefont{Maloney and Lema\^\i{}tre}(2006)}]{maloney06}
\bibinfo{author}{\bibfnamefont{C.~E.} \bibnamefont{Maloney}} \bibnamefont{and}
  \bibinfo{author}{\bibfnamefont{A.}~\bibnamefont{Lema\^\i{}tre}},
  \bibinfo{journal}{Phys. Rev. E} \textbf{\bibinfo{volume}{74}},
  \bibinfo{pages}{016118} (\bibinfo{year}{2006}).

\bibitem[{\citenamefont{Tighe}(2011)}]{tighe11}
\bibinfo{author}{\bibfnamefont{B.~P.} \bibnamefont{Tighe}},
  \bibinfo{journal}{Phys. Rev. Lett.} \textbf{\bibinfo{volume}{107}},
  \bibinfo{pages}{158303} (\bibinfo{year}{2011}).

\bibitem[{\citenamefont{V\aa{}gberg et~al.}(2011)\citenamefont{V\aa{}gberg,
  Olsson, and Teitel}}]{vagberg11}
\bibinfo{author}{\bibfnamefont{D.}~\bibnamefont{V\aa{}gberg}},
  \bibinfo{author}{\bibfnamefont{P.}~\bibnamefont{Olsson}}, \bibnamefont{and}
  \bibinfo{author}{\bibfnamefont{S.}~\bibnamefont{Teitel}},
  \bibinfo{journal}{Phys. Rev. E} \textbf{\bibinfo{volume}{83}},
  \bibinfo{pages}{031307} (\bibinfo{year}{2011}).

\bibitem[{\citenamefont{Wyart}(2005)}]{wyartannales}
\bibinfo{author}{\bibfnamefont{M.}~\bibnamefont{Wyart}},
  \bibinfo{journal}{Annales de Physique} \textbf{\bibinfo{volume}{30}},
  \bibinfo{pages}{1} (\bibinfo{year}{2005}).

\bibitem[{\citenamefont{Schlegel et~al.}(2016)\citenamefont{Schlegel, Brujic,
  Terentjev, and Zaccone}}]{schlegel16}
\bibinfo{author}{\bibfnamefont{M.}~\bibnamefont{Schlegel}},
  \bibinfo{author}{\bibfnamefont{J.}~\bibnamefont{Brujic}},
  \bibinfo{author}{\bibfnamefont{E.}~\bibnamefont{Terentjev}},
  \bibnamefont{and} \bibinfo{author}{\bibfnamefont{A.}~\bibnamefont{Zaccone}},
  \bibinfo{journal}{Scientific Reports} \textbf{\bibinfo{volume}{6}},
  \bibinfo{pages}{18724} (\bibinfo{year}{2016}).

\bibitem[{\citenamefont{Goodrich et~al.}(2012)\citenamefont{Goodrich, Liu, and
  Nagel}}]{goodrich12}
\bibinfo{author}{\bibfnamefont{C.~P.} \bibnamefont{Goodrich}},
  \bibinfo{author}{\bibfnamefont{A.~J.} \bibnamefont{Liu}}, \bibnamefont{and}
  \bibinfo{author}{\bibfnamefont{S.~R.} \bibnamefont{Nagel}},
  \bibinfo{journal}{Phys. Rev. Lett.} \textbf{\bibinfo{volume}{109}},
  \bibinfo{pages}{095704} (\bibinfo{year}{2012}).

\bibitem[{\citenamefont{Dagois-Bohy et~al.}(2012)\citenamefont{Dagois-Bohy,
  Tighe, Simon, Henkes, and van Hecke}}]{dagois-bohy12}
\bibinfo{author}{\bibfnamefont{S.}~\bibnamefont{Dagois-Bohy}},
  \bibinfo{author}{\bibfnamefont{B.~P.} \bibnamefont{Tighe}},
  \bibinfo{author}{\bibfnamefont{J.}~\bibnamefont{Simon}},
  \bibinfo{author}{\bibfnamefont{S.}~\bibnamefont{Henkes}}, \bibnamefont{and}
  \bibinfo{author}{\bibfnamefont{M.}~\bibnamefont{van Hecke}},
  \bibinfo{journal}{Phys. Rev. Lett.} \textbf{\bibinfo{volume}{109}},
  \bibinfo{pages}{095703} (\bibinfo{year}{2012}).

\bibitem[{\citenamefont{Goodrich et~al.}(2014)\citenamefont{Goodrich,
  Dagois-Bohy, Tighe, van Hecke, Liu, and Nagel}}]{goodrich14}
\bibinfo{author}{\bibfnamefont{C.~P.} \bibnamefont{Goodrich}},
  \bibinfo{author}{\bibfnamefont{S.}~\bibnamefont{Dagois-Bohy}},
  \bibinfo{author}{\bibfnamefont{B.~P.} \bibnamefont{Tighe}},
  \bibinfo{author}{\bibfnamefont{M.}~\bibnamefont{van Hecke}},
  \bibinfo{author}{\bibfnamefont{A.~J.} \bibnamefont{Liu}}, \bibnamefont{and}
  \bibinfo{author}{\bibfnamefont{S.~R.} \bibnamefont{Nagel}},
  \bibinfo{journal}{Phys. Rev. E} \textbf{\bibinfo{volume}{90}},
  \bibinfo{pages}{022138} (\bibinfo{year}{2014}).

\bibitem[{\citenamefont{Feng et~al.}(1985)\citenamefont{Feng, Thorpe, and
  Garboczi}}]{feng85}
\bibinfo{author}{\bibfnamefont{S.}~\bibnamefont{Feng}},
  \bibinfo{author}{\bibfnamefont{M.}~\bibnamefont{Thorpe}}, \bibnamefont{and}
  \bibinfo{author}{\bibfnamefont{E.}~\bibnamefont{Garboczi}},
  \bibinfo{journal}{Physical Review B} \textbf{\bibinfo{volume}{31}},
  \bibinfo{pages}{276} (\bibinfo{year}{1985}).

\bibitem[{\citenamefont{Merkel et~al.}(2019)\citenamefont{Merkel, Baumgarten,
  Tighe, and Manning}}]{merkel19}
\bibinfo{author}{\bibfnamefont{M.}~\bibnamefont{Merkel}},
  \bibinfo{author}{\bibfnamefont{K.}~\bibnamefont{Baumgarten}},
  \bibinfo{author}{\bibfnamefont{B.~P.} \bibnamefont{Tighe}}, \bibnamefont{and}
  \bibinfo{author}{\bibfnamefont{M.~L.} \bibnamefont{Manning}},
  \bibinfo{journal}{Proc. Nat. Acad. Sci. (USA)}
  \textbf{\bibinfo{volume}{116}}, \bibinfo{pages}{6560} (\bibinfo{year}{2019}).

\bibitem[{\citenamefont{Ellenbroek et~al.}(2009)\citenamefont{Ellenbroek,
  Zeravcic, van Saarloos, and van Hecke}}]{ellenbroek09b}
\bibinfo{author}{\bibfnamefont{W.~G.} \bibnamefont{Ellenbroek}},
  \bibinfo{author}{\bibfnamefont{Z.}~\bibnamefont{Zeravcic}},
  \bibinfo{author}{\bibfnamefont{W.}~\bibnamefont{van Saarloos}},
  \bibnamefont{and} \bibinfo{author}{\bibfnamefont{M.}~\bibnamefont{van
  Hecke}}, \bibinfo{journal}{EPL} \textbf{\bibinfo{volume}{87}},
  \bibinfo{pages}{34004} (\bibinfo{year}{2009}).

\bibitem[{\citenamefont{Baumgarten and Tighe}(2017)}]{baumgarten17b}
\bibinfo{author}{\bibfnamefont{K.}~\bibnamefont{Baumgarten}} \bibnamefont{and}
  \bibinfo{author}{\bibfnamefont{B.~P.} \bibnamefont{Tighe}},
  \bibinfo{journal}{Soft Matter} \textbf{\bibinfo{volume}{13}},
  \bibinfo{pages}{8368 } (\bibinfo{year}{2017}).

\bibitem[{\citenamefont{Dagois-Bohy et~al.}(2017)\citenamefont{Dagois-Bohy,
  Somfai, Tighe, and van Hecke}}]{dagois-bohy17}
\bibinfo{author}{\bibfnamefont{S.}~\bibnamefont{Dagois-Bohy}},
  \bibinfo{author}{\bibfnamefont{E.}~\bibnamefont{Somfai}},
  \bibinfo{author}{\bibfnamefont{B.}~\bibnamefont{Tighe}}, \bibnamefont{and}
  \bibinfo{author}{\bibfnamefont{M.}~\bibnamefont{van Hecke}},
  \bibinfo{journal}{Soft Matter} \textbf{\bibinfo{volume}{13}},
  \bibinfo{pages}{9036 } (\bibinfo{year}{2017}).

\end{thebibliography}

\end{document}